\documentclass[prl,twocolumn,english,showpacs]{revtex4-1}
\usepackage{graphicx}

\usepackage{amsmath,amsfonts,amssymb}	
\usepackage{mathrsfs}					

\usepackage{microtype}
\usepackage{bm}

\begin{document}

\title{Two-dimensional Fermi liquid with attractive interactions}

\author{B. Fr{\"o}hlich$^1$}
\author{M. Feld$^1$}
\author{E. Vogt$^1$}
\author{M. Koschorreck$^1$}
\author{M. K{\"o}hl$^1$}
\author{C. Berthod$^2$}
\author{T. Giamarchi$^2$}
\affiliation{$^1$Cavendish Laboratory, University of Cambridge, JJ Thomson Avenue, Cambridge CB3 0HE, United Kingdom\\$^2$DPMC-MaNEP, University of Geneva, 24 Quai Ernest-Ansermet, CH-1211 Geneva, Switzerland}

%

\begin{abstract}
We realize and study an attractively interacting two-dimensional Fermi liquid. Using momentum-resolved photoemission spectroscopy we measure the self-energy, determine the contact parameter of the short-range interaction potential, and find their dependence on the interaction strength. We successfully compare the measurements to a theoretical analysis taking properly into account finite temperature, the harmonic trap, and the averaging over several two-dimensional gases with different peak densities.
\end{abstract}

\maketitle

Atomic quantum gases have been proposed as quantum simulators to identify the microscopic origin of condensed matter phenomena which have been pondered for decades. One such phenomenon is the Fermi liquid \cite{Landau1956}, that has been the cornerstone of the description of solids for the last fifty years. The underlying concept of this remarkable theory is that, although the basic quantum particles can be strongly interacting, there are some excitations -- named Landau quasiparticles -- which are essentially non-interacting. These excitations have the same quantum numbers (charge and spin) as the original particles, but their dynamical properties can be significantly different. The quasiparticle dynamics is described by a fundamental function called the self-energy, whose real and imaginary parts encode information about the quasiparticle dispersion and decay, respectively. In general, this function has a rich structure representing the variety of single-particle excitations as a function of momentum and energy. In the low-energy and low-temperature regime, the self-energy reduces to essentially two numbers: the quasiparticle effective mass and the life-time. For interacting Fermi gases, the quasiparticle picture allows therefore to summarize the effects of all interactions in the redefinition of these two parameters and to treat the interacting quantum system as a free-fermion gas of quasiparticles. This constitutes an immense simplification over a fully interacting quantum system and has often been the starting point to understand more complex phenomena such as semiconductors and the transistor, superconductivity and the BCS theory, and, more recently, the giant magnetoresistance.

Initial experimental verifications of Fermi liquid theory were quite indirect and mostly based on collective mode propagation and transport measurements, for example in simple metals, heavy-fermion materials (CeAl$_3$, LaRu$_2$Si$_2$), and liquid $^3$He. Only in recent years the Angle Resolved Photoemission Spectroscopy (ARPES) technique has allowed for a direct measurement for the probability to find a single-particle excitation with a given momentum and energy -- the so-called spectral function \cite{Damascelli2003}. It has therefore been instrumental in providing a direct measure for the existence and properties of quasiparticles. However, the analysis of APRES spectra and a detailed comparison with the theory remains difficult in solid-state systems, due to non-trivial interactions in the final state, and insufficient knowledge of the dispersion even for the non-interacting particles. The clean Fermi liquid system of $^3$He does not easily lend itself to the equivalent of an ARPES measurement.

Experiments with cold atomic gases provide a remarkable alternative to tackle the question of interactions in quantum fluids. These systems have the advantage of combining short range interactions with an unprecedented control of the interaction strength. They also offer control of the dimensionality of the system, and in particular have allowed the realization of interacting two-dimensional fermionic systems \cite{Frohlich2011,Dyke2011,Feld2011,Sommer2012}. It was thus natural to develop the equivalent of the ARPES technique for cold atoms to probe the quasiparticle dynamics and the characteristics and properties of the Fermi liquid \cite{Dao2007}. However, despite successes in realizing ARPES experiments \cite{Stewart2008,Feld2011,Koschorreck2012} and probing the formation of a gap for attractive interactions, no comparison with the properties of two-dimensional Fermi liquids has yet been made. Some aspects of Fermi liquid properties in cold gases have so far been probed only in three dimensions \cite{Nascimbene2011,Sommer2011} by studying the magnetic susceptibility in the strongly interacting regime above $T_c$.

\begin{figure}
\includegraphics[width=\columnwidth,clip=true]{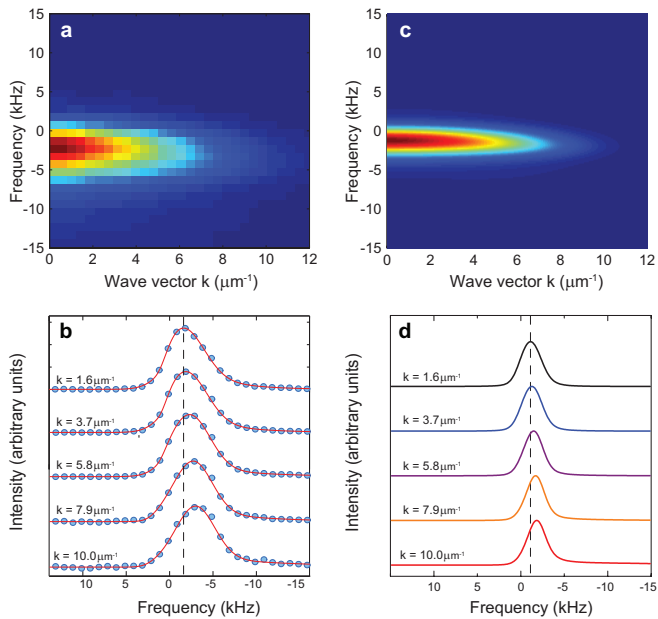}
  \caption{(Color online) Momentum-resolved photoemission signal in experiment and theory. \textbf{a)} Measured signal at $1/\ln(k_Fa_{2D})=0.35$. A free particle would correspond to a straight horizontal line. \textbf{b)} Energy distribution curves at different momenta $\hbar k$. The solid lines represent a fit to the data. We see a shift of the maximum towards lower energies which indicates the effective mass of the quasi particles. The dashed line indicates $E_0$, the position of the peak at $k=0$. \textbf{c)} and \textbf{d)} Calculated intensity and energy distribution curves for the same parameters as in the experiment.}
 \label{fig1}
\end{figure}

Here, using momentum-resolved radiofrequency (r.f.) spectroscopy \cite{Dao2007,Stewart2008,Feld2011,Koschorreck2012}, we extract the self-energy of a two-dimensional Fermi gas with attractive interactions. We find quantitative agreement with calculations based on Fermi liquid theory. Moreover, we show that the Hartree energy term can play a dominant role in the quantitative understanding of ARPES spectra in harmonically confined Fermi gases.

We prepare a quantum degenerate Fermi gas of $^{40}$K atoms in the $|F=9/2,m_F=-9/2\rangle\equiv |-9/2\rangle$ and $|F=9/2,m_F=-7/2\rangle\equiv |-7/2\rangle$ hyperfine states in a one-dimensional optical lattice of wavelength $\lambda=1064$\,nm, populating a stack of approximately 40 individual two-dimensional quantum gases \cite{Frohlich2011}. In the central layers we confine a few thousand atoms per two-dimensional gas. The radial confinement of the two-dimensional gases is harmonic with a trap frequency of $\omega_r=2\pi\times 127$\,Hz and the axial trap frequency is $\omega_z=2 \pi \times 75$\,kHz. After we ramp to the desired magnetic field value near the Feshbach resonance between the  $|-9/2\rangle$ and $|-7/2\rangle$ states, we perform momentum-resolved radiofrequency spectroscopy between the $|-7/2\rangle$ and $|F=9/2,m_F=-5/2\rangle\equiv|-5/2\rangle$ states \cite{Feld2011,Koschorreck2012}. To this end, we apply a radio frequency (r.f.) pulse of approximately $\Omega_{\textrm{r.f.}}=47$\,MHz with a Gaussian amplitude envelope with a full width at half maximum of 280\,$\mu$s. After a further 100\,$\mu$s we turn off the optical lattice, switch off the magnetic field and separate the three spin components by applying a magnetic field gradient. After a free expansion of the gas of 12\,ms we take an absorption image and average the density distribution of the $|-5/2\rangle$ component azimuthally to obtain the density $n(\nu,k=\sqrt{k_x^2+k_y^2})$, where we have taken the energy zero at the Zeeman energy $E_Z(B)$ of the spin-flip transition in vacuum, i.e., $\nu=E_Z(B)/h-\Omega_\textrm{r.f.}$.

The parameter $g$ characterizing the interaction strength of a two-dimensional Fermi liquid is given by $g=-2 \pi \hbar^2/[m\ln(k_Fa_{2D})]$ \cite{Bloom1975}. Here, $E_F$ is the Fermi energy, $k_F$ is the Fermi wave vector, and $a_{2D}$ is the two-dimensional scattering length defined via the binding energy of the confinement-induced bound state $E_B=\hbar^2/m a_{2D}^2$. On the attractive side of a Feshbach resonance $E_B$ becomes exponentially small, justifying the role of $g$ as a small parameter. We consider the weakly interacting regime $0\leq 1/\ln(k_Fa_{2D})<1$ and study it in the normal state where the thermal energy $k_BT$ is much larger than the energy scale for pairing. This realizes a Fermi liquid state without complications of (pseudogap) pairing. In Fig. 1a we show a  momentum-resolved photoemission signal for $1/\ln(k_Fa_{2D})=0.35$ at $T/T_{\rm F}=0.27$, where the free-particle dispersion has been implicitly subtracted. Here, $T$ is the temperature and $T_F=E_F/k_B$. Figure 1b shows the corresponding energy distribution curves (EDC) for different values of the wave vector $k$. In order to take into account the slightly asymmetric shape of the peak, we use a combination of a Gaussian and a modified Gumbel distribution to fit the energy distribution curves \cite{Feld2011}, which we find to capture the feature very well for all data sets taken. We compare our experimental data to a theoretical calculation using the ladder approximation \cite{Engelbrecht1992} parameterized by the bound-state energy $E_B$, which takes the experimental conditions, such as inhomogeneity due to the trap and finite temperature, fully into account (see Supplementary Material). Our theoretical modeling improves over previous work, which has focussed on homogeneous Fermi liquids at zero temperature with repulsive \cite{Bloom1975,Engelbrecht1992} and attractive \cite{Miyake1983,Nozieres1985,Randeria1989} interactions, as well as on repulsive interactions at finite temperature \cite{Chitov2001}. The result of our calculation for $1/\ln(k_Fa_{2D})=0.35$ is displayed in Fig.~1c and d.  An energy-resolution broadening of 1.5 kHz was applied to the theoretical data, which is experimentally measured for the non-interacting gas and which corresponds to the Fourier-limited width of the r.f. pulse. For the interacting gas, we experimentally observe a larger width, which is not capture by theory, and which therefore possibly stems from final state interactions.

\begin{figure}
\includegraphics[width=0.8\columnwidth,clip=true]{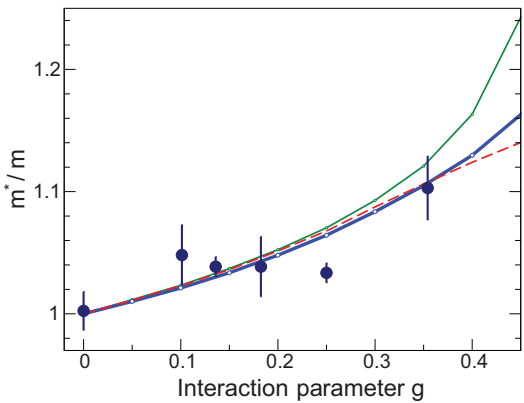}
  \caption{(Color online) Dependence of the effective mass parameter on the interaction parameter $1/\ln(k_Fa_{2D})$. The blue solid line shows the numerical calculation of the trap average and averaging over a density distribution with an rms width of 46 layers for the experimental conditions. The red dashed curve is the trap average for the central layer only. The green solid curve includes only the Hartree term  in the self energy.}
 \label{fig2}
\end{figure}

We analyze the dispersion of the peak in Fig. 1 by means of two parameters, which are the $k=0$ intercept $E_0$, and the curvature represented by an effective-mass parameter $m^*$, according to $E_{max}(k)=E_0+\frac{\hbar^2k^2}{2}\left(\frac{1}{m^*}-\frac{1}{m}\right)$. In Fig. 2 we show the effective mass parameter for different values of $1/\ln(k_Fa_{2D})$ at $T/T_{\rm F}=0.27$. For zero interaction ($g = 0$), $m^*$ equals the free-particle mass to within $1\%$ and $E_0=-1.0(0.3)$\,kHz. This data point calibrates our weak final state interactions. Increasing the interaction strength on the attractive side of the Feshbach resonance leads to an increase of $m^*$ as the dressing of the bare fermions increases. Experimentally, $E_0$ does not show a significant variation over this range.

Our experimental results and theoretical calculations display very good agreement with each other (see Fig. 2). The solid blue line shows the theoretical $m^*/m$ for the experimental Fermi energy and temperature and averaged over a distribution of 2D gases with a Gaussian envelope of Fermi energies with a r.m.s. width of 46 layers. The dashed red line shows the result for the trap average of the center layer only, indicating that the averaging over several layers only has a very minor effect. When only the first-order Hartree term in the self-energy, which is a density-dependent shift without dynamical consequences, is taken into account (green curve in Fig. 2), one nevertheless obtains a non-linear contribution to the dispersion, which is a result of the inhomogeneity of the system. This shows that the Hartree term \cite{Chen2009,Pieri2011,Perali2011} plays an important role in the quantitative interpretation of momentum-resolved r.f. spectra in confined geometries. The higher-order dynamical corrections reduce the dispersion associated with the Hartree term. For $1/\ln(k_Fa_{2D})<0.4$, the effect of these corrections is at the limit of the experimental error bars. In order to study the temperature dependence of the Fermi liquid properties, we vary the temperature at approximately constant $1/\ln(k_Fa_{2D})$. We observe that $m^*$ decreases with increasing temperature and approaches the bare particle mass $m^*/m\approx 1$ at approximately $T=T_F$ (see Figure~3), which agrees very well with theory.
\begin{figure}
\includegraphics[width=.8\columnwidth,clip=true]{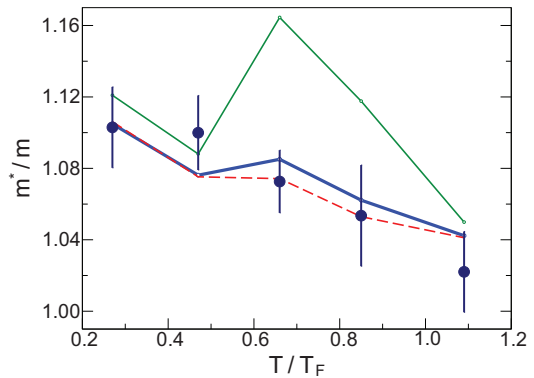}
  \caption{(Color online) Temperature dependence of the effective mass parameter. For our lowest temperature $T/T_F=0.27$ this corresponds to an interaction strength of $1/\ln(k_Fa_{2D})=0.35$; samples at higher temperature contain more atoms and have a slightly larger Fermi energy, which decreases the coupling strength to $1/\ln(k_Fa_{2D})=0.32$ at $T/T_F=1.09$. The lines show the numerical simulation for the experimental parameters. The blue line is the trap average over 46 layers corresponding to the experimental conditions. The red dashed curve is the trap average for the central layer only. The green line includes the Hartree term only in the self energy. The difference between Hartree-only and the full self-energy changes with the number of trapped atoms, showing the intricate relation between the Hartree energy and the effective mass parameter.}
 \label{fig3}
\end{figure}

Finally, we turn our attention to the contact parameter $C$ \cite{Tan2008a,Combescot2009,Werner2010,Stewart2010,Kuhnle2011,Langmack2012}. The contact is determined by the two-particle correlation function  between atoms of opposite spin at short distance $g^{(2)}_{\uparrow \downarrow}(|r_\uparrow -r_\downarrow|)$, and it governs the momentum distribution of a gas at large momenta according to $n(k)\sim C/k^4$ for $k\gg k_F$. Through the universal Tan relations \cite{Tan2008a}, the contact parameter provides an important link between the microscopic physics of the short-range atom-atom interactions and thermodynamic quantities. In three dimensions, the contact has been measured from photoemission spectra \cite{Stewart2010} and Bragg scattering \cite{Kuhnle2011}, and Tan's relations have been experimentally verified. In two dimensions, Tan's relations have to be refined, and the spectral line shape of r.f. spectra receives non-trivial corrections \cite{Langmack2012}.

The dimensionless contact parameter $C^\prime=C/k_{\rm F}^2$ can be measured from the high-frequency tail of the momentum-integrated single-particle spectral function \cite{Schneider2010b}. In two dimensions, the spectral intensity $I_\sigma(\nu^\prime)$ normalized to the intensity of the r.f. pulse relates to the contact by \cite{Langmack2012}
\begin{equation}
I_\sigma(\nu^\prime) = \frac{C^\prime}{2\pi\nu^{\prime 2} }\frac{\ln^2(\tilde{E}_B/E_B)}{\ln^2(\nu^\prime E_F/\tilde{E}_B)+\pi^2}=C^\prime\times \cal{I}(\nu^\prime). \label{eq:I}
\end{equation}
Here, $\nu^\prime=h\nu/E_F$ and $\tilde{E}_B=\tilde{E}_{B,3D}$ is the binding energy of the most weakly bound state of atoms in the final state. Since the three-dimensional binding energy between the $|-9/2\rangle$ and $|-5/2\rangle$ states is $\tilde{E}_{B,3D}\approx h\times3$\,MHz \cite{Chin2010} (in the relevant magnetic field range between 204\,G and 209\,G) and therefore much larger than $\hbar \omega_z$, the effects of quasi-two-dimensional confinement on the binding energy can be neglected. Generally, final state interactions can play a significant role for the contact in two dimensions since their contribution disappears only logarithmically with increasing binding energy of the final state.

We extract the contact $C^\prime$ from the data by dividing the momentum-integrated intensity of the spectrum by the function $\cal{I}(\nu^\prime)$ and fitting the resulting constant at large $\nu^\prime$. The inset in Figure~4 shows an example for a typical data set. In Fig. 4 we plot the measured $C^\prime$ as a function of $1/\ln(k_Fa_{2D})$ at $T/T_F=0.27$ (solid blue points). We compare our experimental results with the theory for the trapped gas at finite temperature (solid blue line). The contact was calculated using the theoretical momentum-integrated spectral function and fitting a pure $1/\nu^2$ decay as the theory does not include final state interactions. Our calculations show that the value of $C'$ is reduced by temperature and by the inhomogeneity, due to a transfer of spectral weight induced by the Hartree term from the $1/\nu^2$ tail to low energy.

The agreement is excellent in the regime of weak coupling. The calculation of the temperature-dependent contact at larger coupling requires further work, as the effects of the bound state become important when $E_B$ approaches $k_BT$. For comparison, we also show the zero-temperature prediction of the contact for a homogeneous system based on a quantum-Monte-Carlo calculation \cite{Bertaina2011} (dashed gray line). In order to derive the contact from the total energy data of reference \cite{Bertaina2011}, we have used the adiabatic theorem $dE'/d[\ln(k_{\rm F}a_{\rm 2D})]=C'/\pi$, where $E^\prime=E/E_F$. In the weak coupling regime, both the experimental results as well as our theoretical values are slightly below the zero-temperature theory. This is the expected behaviour of the contact, which decreases with increasing temperature. In the strongly interacting regime, our data come closer to the zero temperature prediction, possibly because when $E_B>k_BT$ the contribution of the bound state to the contact becomes more dominant. Finally, we also show the prediction of the contact for the homogeneous Fermi liquid at zero temperature, which has been derived from the power series expansion of the total energy per particle \cite{Engelbrecht1992} $2E^\prime/N=1-1/\ln(k_Fa_{2D})+A/\ln(k_Fa_{2D})^2+...$ with $A=3/4-\ln(2)$.

The above results show that momentum-resolved r.f. spectroscopy can be employed to extract important information about a Fermi liquid, such as the self-energy, and that disentangling the dynamical part of the self-energy from the non-trivial contributions arising from the Hartree term in the trap is important. The latter could only be overcome using confining potentials different from the usually employed harmonic potential, since our theoretical analysis shows that the Hartree contribution is independent of the strength of the harmonic potential.
\begin{figure}
\includegraphics[width=.8\columnwidth,clip=true]{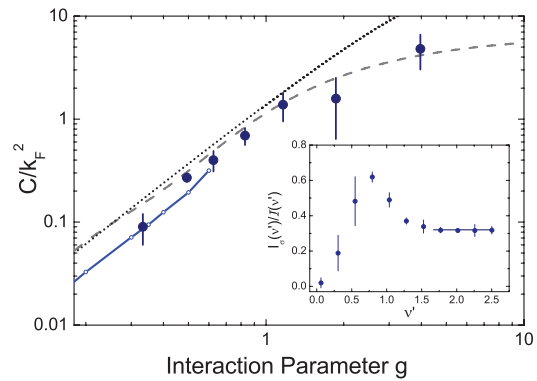}
  \caption{(Color online) Dimensionless contact $C'=C/k_{\rm F}^2$ in the Fermi liquid regime. We compare with our theoretical results for the trap average in the weakly interacting regime (solid blue line). Moreover, we show the quantum Monte-Carlo calculation at zero temperature \cite{Bertaina2011} (dashed gray line) and the second-order Fermi liquid prediction (dotted line). Inset: To determine the contact we divide the measured r.f. transition rate $I_\sigma(\nu)$ by $\cal{I}(\nu)$ (see text) and fit a constant at large $\nu$. This data set is for $1/\ln(k_Fa_{2D})=0.49$.}
 \label{fig4}
\end{figure}

We thank C. Kollath, D. Pertot, and W. Zwerger for discussions. The work has been supported by {EPSRC} (EP/G029547/1, EP/J01494X/1), Daimler-Benz Foundation (B.F.), Studienstiftung, and DAAD (M.F.), Swiss NSF under MaNEP and Division II.

\end{document}